# Investigation of the longitudinal component of an electron electromagnetic field under condition of the shadowing effect

G.-A. Naumenko, L.-G. Sukhikh, Yu.-A. Popov, M.-V. Shevelev

Physical and technical institute. Tomsk polytechnic university. 634050 Lenina Ave 2a, Tomsk, Russia.

In this paper we present the method and experimental results of the investigation of a longitudinal component of relativistic electron electromagnetic field in the shadow area of a transversal component. We show experimentally, that in a region, comparable with the formation length area no shadowing effect of the longitudinal component of relativistic electron electromagnetic field appears. This is important for understanding of possibility of the shadowing effect in Smith-Purcell radiation and some other radiation types.

## 1. - Introduction

In principle the radiation emitted by relativistic electrons passing through or near material targets can be calculated using the macroscopic Maxwell equations. However so far such calculations for thick conductive targets, when the thickness is much larger than a skin-layer, are absent. Therefore phenomenological concepts, like "surface current viewpoint" and "pseudo-photons", are widely used for calculation of transition and diffraction radiation from conductive targets. Furthermore these concepts are useful for the intuitive understanding of the main features of these phenomena. In [1] was shown, that in contrast to the "pseudo-photon" method the "surface current viewpoint" is not applicable for a forward transition and diffraction radiation explanation.

The shadowing effect is the effect when an electron loses a part of its Coulomb field (see also "semi-bare electron" in [2]). This effect was investigated experimentally in [3] in macroscopic mode. In this paper we had shown that electron field is suppressed just downstream to a conductive or absorbing screen and this field is recovered with distance from screen. In [1] was shown that the shadowing effect may be explained in frame of pseudo-photon point of view [4,5]. However this method may be applied only for a transversal component of an electromagnetic field of relativistic electron and not applicable for a longitudinal one.

On the other hand, the longitudinal component of an electron electromagnetic field play the basic role in such processes as the Smith-Purcell radiation, diffraction radiation at grazing angles and so on. Therefore the knowledge about properties of the longitudinal component of an electron electromagnetic field under condition of the shadowing effect is very important for understanding of the nature of these phenomena. This work is devoted to an experimental investigation of this problem.

## 2. - Methodical basis

Electric field of relativistic electron has an axial symmetry and may be presented in Fourier approximation by expression (1)

$$\vec{E}_e(\vec{\rho}, z, \lambda) = \begin{Bmatrix} \vec{E}_\perp \\ E_\parallel \end{Bmatrix} = \frac{2e}{\gamma \lambda \beta^2} e^{i\frac{2\pi}{\lambda}z} \begin{Bmatrix} \frac{\vec{\rho}}{\rho} K_1\left(\frac{2\pi}{\gamma \lambda \beta}\rho\right) \\ -\frac{i}{\gamma} K_0\left(\frac{2\pi}{\gamma \lambda \beta}\rho\right) \end{Bmatrix},$$ (1)

where $\vec{E}_e$ is the electron electric strength vector, z is the coordinate of electron in direction of the electron motion in respect to a observation point, $\vec{\rho}$ is vector of observation point transversal coordinates, $E_\parallel$ is the longitudinal component of the electron electric strength vector and $\vec{E}_\perp$ is the transversal one, $\lambda$ is the wavelength ($\lambda = \frac{2\pi}{\omega}$, $\omega$ is the Fourier approximation variable), $e$ is the electron charge, $\gamma$ is the Lorenz-factor, $\beta$ is the electron velocity (light velocity is assumed to be equal 1), $K_0$ and $K_1$ are Bessel functions.

We can see that relation $\frac{E_\parallel^2}{E_\perp^2} \approx \frac{1}{\gamma^2}$ for relativistic electron is very small (for $\gamma = 12$ $\frac{1}{\gamma^2} = \frac{1}{144}$). It seems that it is impossible to separate a longitudinal component over a transversal one experimentally.

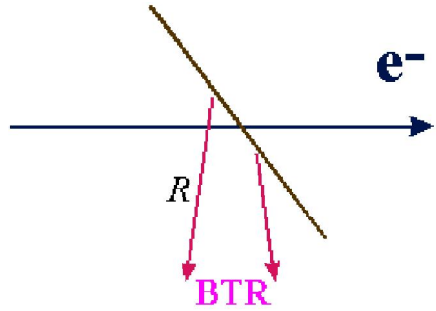

Fig. 1. - Scheme of backward transition radiation

However, let us consider next simple scheme (Fig. 1). An electric strength vector of backward transition radiation (BTR) from conductive target (or backward diffraction radiation (BDR) of electron moving through a small hole) may be presented using Kirhoff integral (2) (see [6]).

$$\vec{E}(\vec{r}) = 2\int_{S_1} (\vec{n} \times \vec{E}_e) \times grad'G \, da',$$ (2)

where $G(\vec{r}, \vec{r}') = \frac{1}{4\pi} \cdot \frac{e^{ikR}}{R}$ is Green function of a target surface, $\vec{n}$ is the vector perpendicular to the target surface normalized to the unit. Using (2) and (1) we may calculate the angular distribution of radiation intensity $|\vec{E}|^2$ in radiation plane for $\gamma = 12$ in far field zone as a function of the observation angle $\theta$ (see case *a* in Fig. 2). We can see in this picture an asymmetry of the angular distribution. In order to clarify a cause of this asymmetry we neglect the longitudinal component $E_\parallel$. In this case (Fig. 2 *b*) asymmetry disappears. In opposite case if we suppress the transversal component $E_\perp$ (see Fig. 2*c*), the asymmetry is enhanced markedly. I.e. the BTR (or BDR) asymmetry may be used as a sensitive tool for experimental investigation of the longitudinal component of an electron electric field strength.

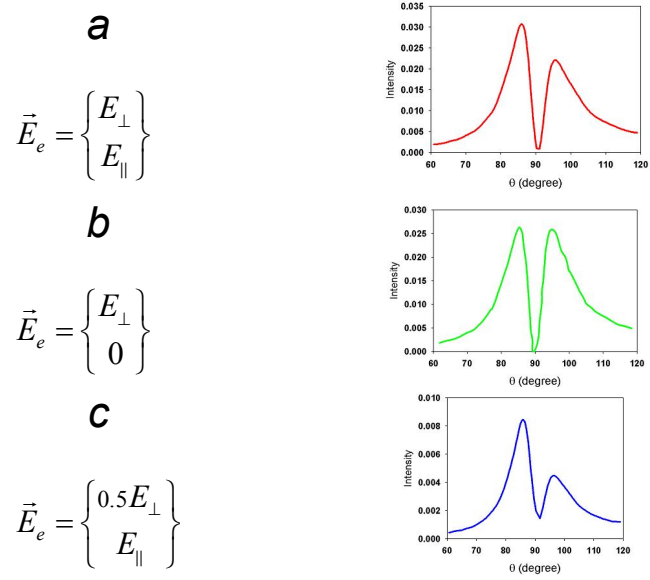

*a*

$$\vec{E}_e = \begin{Bmatrix} E_\perp \\ E_\parallel \end{Bmatrix}$$

*b*

$$\vec{E}_e = \begin{Bmatrix} E_\perp \\ 0 \end{Bmatrix}$$

*c*

$$\vec{E}_e = \begin{Bmatrix} 0.5 E_\perp \\ E_\parallel \end{Bmatrix}$$

Fig. 2. - BTR asymmetry.

3. - Experiment

One may argue that the radiation intensity in this wavelength region during the interaction of the electron field with the targets is negligible and not accessible for a measurement. However, this is not the case because of the coherent character of radiation. Actually if the number of electron in a bunch Ne>>1, the radiation intensity from the bunch may be presented as

$$I_b \approx N_e^2 \cdot f^2\left(\frac{\lambda}{\sigma}\right) \cdot I_e,$$

where $f\left(\frac{\lambda}{\sigma}\right)$ is a form-factor of electron bunch, $\sigma$ is the r.m.s. bunch length. For $\lambda >> \sigma$  $f\left(\frac{\lambda}{\sigma}\right) \approx 1$ and $I_b \approx N_e^2 \cdot I_e$ instead $I_b \approx N_e \cdot I_e$, which take place for incoherent radiation. For the electron beam of microtron at Tomsk Nuclear Physics Institute with the parameters showing in Table 1 the coherent radiation intensity for $\lambda$>9 mm is by 8 orders larger than incoherent one and has the power level $\approx$1 Watt per steradian. It means one may investigate coherent radiation in this wavelength range without a problem.

Table 1. - Electron beam parameters.

| Electron energy | 6.1 MeV ($\gamma = 12$). | Bunch period | 380 psec |
|---|---|---|---|
| Train duration | $\tau \approx 4$ $\mu$sec | Bunch population | $N_e$=6·10$^8$ |
| Bunches in a train | $n_b \approx 1.6 \cdot 10^4$ | Bunch length | $\sigma \approx 2$ mm |

The experiment was performed on the extracted electron beam of microtron of Tomsk Nuclear Physics Institute with parameters, shown in table 1. The electron beam is extracted from the vacuum chamber through the beryllium foil with the thickness of 40 $\mu$m.

For the radiation measurements we used the room temperature detector DP20M, with parameters described in [7]. Main elements of the detector are the low-threshold diode, broadband antenna and preamplifier. The detector efficiency in the wavelength region $\lambda$=3~16 mm is estimated as a constant with accuracy ± 15%. The detector sensitivity is 0.3 V/mWatt. The beyond-cutoff wave-guide with $\lambda_{cut}$=17 mm was used to cut the accelerator RF system long wave background. The high frequency limit of wavelength interval is defined by bunch form-factor. This limit ($\lambda_{min} \approx$9 mm) was measured using discrete wave filters [8] and the spectrometer type of grating.

For measurement of the radiation angular distribution asymmetry we used the scheme sown in Fig. 3. The electron beam moves through the hole in absorbing screen.

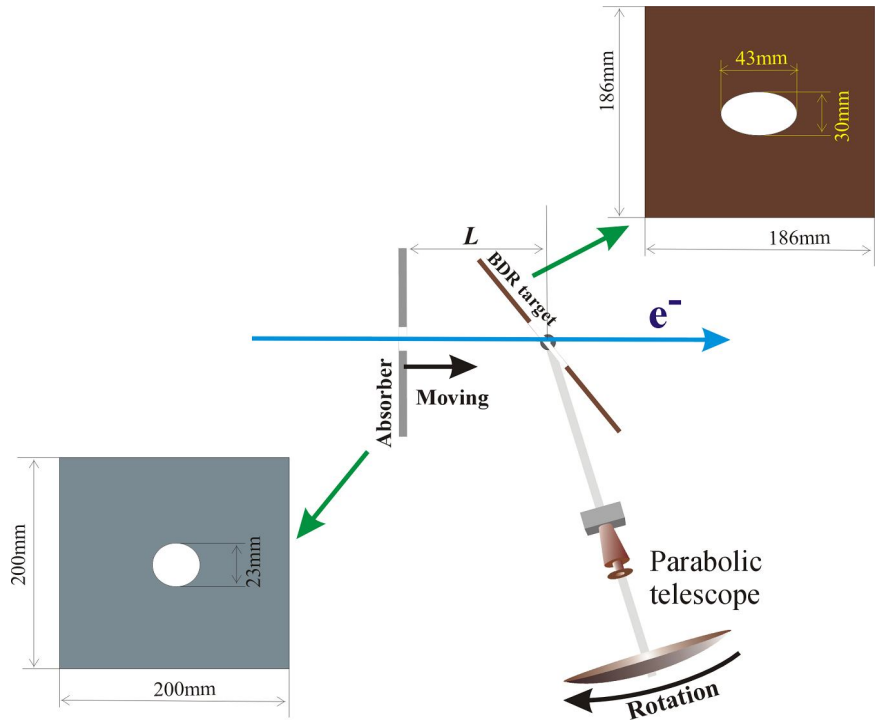

Fig. 3. - Scheme of experiment.

The pseudo-photons of electron field, after shadowing by the screen, are reflected by thick conductive mirror with a hole for electron beam. Last process is a BDR. The mirror is pointed at the angle 45° to the electron beam direction. The minimum value of distance L is limited by the screen and mirror geometry. To exclude the prewave zone effect contribution (see [9]) the parabolic telescope was used for a BDR angular distribution measurement. This method was suggested and tested in [10] and allows us to measure an angular distribution coinciding with one in far field zone (at a distance >> $\gamma^2\lambda$). To exclude the transversal beam size contribution in BDR, the position of the conductive target was fixed and the distance L was varied by the varying of the absorber position.

The beam divergence caused by beryllium window ($\approx$0.08 radian) limits the achievable distance between the absorber and a conductive target by the value $L_{max}\approx$220 mm because of an enhancement of the transversal beam size for large value of L. The measurements were performed with step 1 degree in angular distribution and with step 20 mm in distance L. The statistical error of measured radiation intensity is $\sigma \approx 5\%$. In Fig. 4 is shown the smoothed measured dependence of the radiation intensity on an observation angle $\theta$ and distance L.

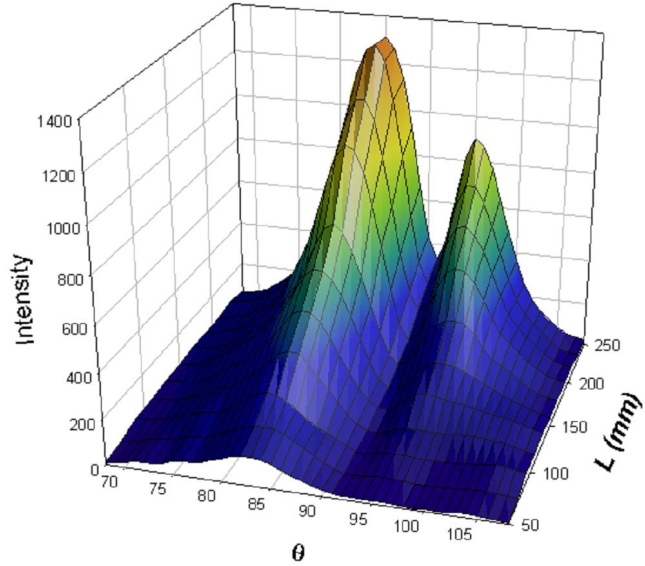

Fig. 4. - Smoothed experimental dependence of BDR.

Using this dependence we may obtain an asymmetry $\eta = \dfrac{M_1 - M_2}{M_1 + M_2}$ as a function of distance $L$ (see Fig. 5), where $M_1$ and $M_2$ are the values of the radiation intensity in the left and right maximum of an angular distribution for a fixed value of $L$.

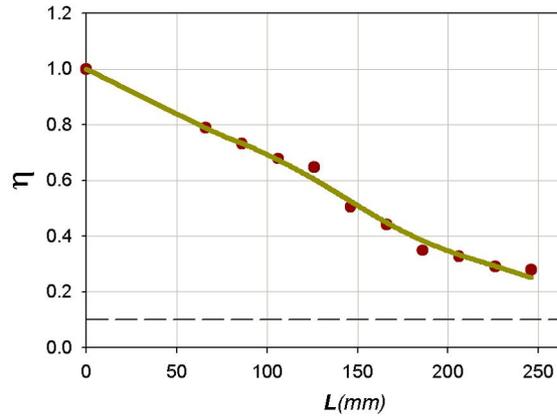

Fig. 5. - Asimmetry as a function of distance $L$.

As was shown above, due to $E_\parallel^2 \ll E_\perp^2$ the main contribution in the average radiation intensity (($M_1+M_2$)/2) is provided by transversal component of an electron field. Therefore we may expect, that the shadowing effect depend mainly on this component. To check this supposition we chose next model. Let us introduce in (1) the suppression factor $\alpha$, which depends on $L$ (see (3)).

$$\vec{E}_e(\vec{\rho}, z, \lambda) = \begin{Bmatrix} E_\perp \\ E_\parallel \end{Bmatrix} = \frac{2e}{\gamma \lambda \beta^2} e^{i\frac{2\pi}{\lambda}z} \begin{Bmatrix} \alpha \dfrac{\vec{\rho}}{\rho} K_1\left(\dfrac{2\pi}{\gamma \lambda \beta}\rho\right) \\ -\dfrac{i}{\gamma} K_0\left(\dfrac{2\pi}{\gamma \lambda \beta}\rho\right) \end{Bmatrix} \quad (3)$$

From (3) for $E_\parallel^2 \ll E_\perp^2$ (as was shown above),

$$(M_1 + M_2)/2 \sim q\alpha^2, \quad (4)$$

where $q$ is proportional factor. Using (3) we may calculate asymmetry as a function of $\alpha^2$ (solid line in Fig. 6). On the other hand, using data from dependence shown in Fig. 4 and taking into account (4) we can obtain the experimental dependence of the measured asymmetry on $\alpha^2$ with accuracy of proportional factor $q$, which is undefined. We may found this factor performing the fit of the experimental data to theoretical dependence by factor $q$. The doted line in Fig. 6 is the fit of the experimental data to theoretical dependence.

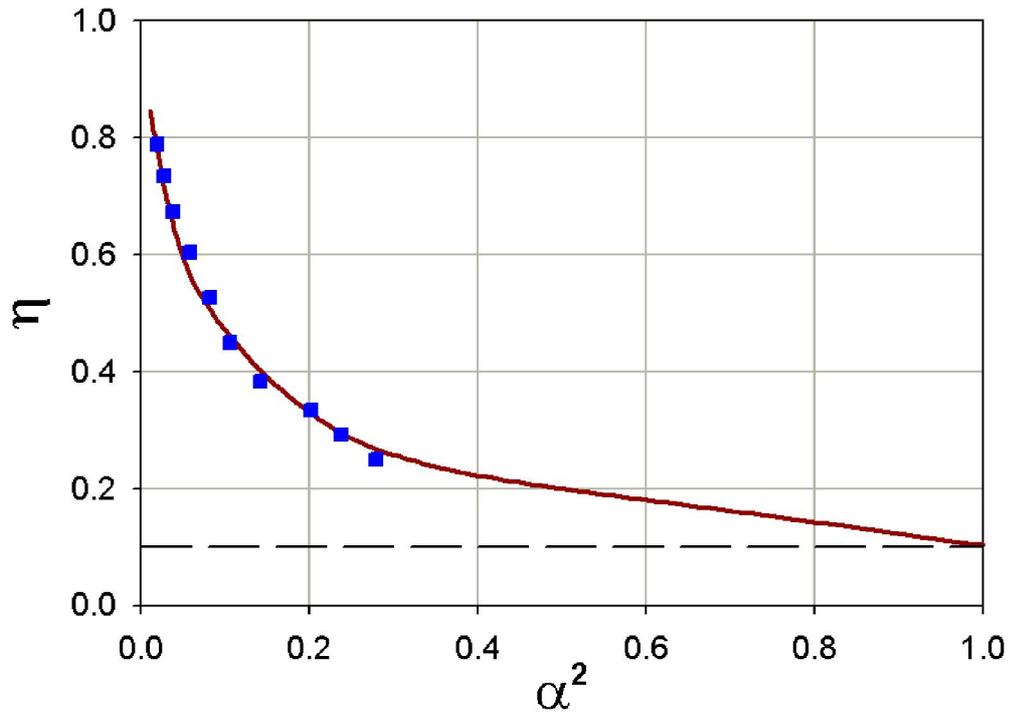

Fig. 6. - Dependence of the asymmetry on the suppression factor $\alpha$. Dots are the experimental points. Solid line is the fit of the theoretical dependence to experimental data.

We can see a good agreement between theoretical calculation and experimental results. It is therefore concluded that the model where the longitudinal component of relativistic electron field does not depend on distance between screen and BDR target is in a good agreement with experiment. In contrast the dashed line in Fig. 6 corresponds to the case when transversal and longitudinal component of electron field are shadowed proportionally.
From the above reasoning it is clear that longitudinal component of electron field does not shadowed in interactions with absorbing screen.

Acknowledgment

This work was partly supported by the warrant-order 1.226.08 of the Ministry of Education and Science of the Russian Federation and by the Federal agency for science and innovation, contract 02.740.11.0245